\documentclass[twocolumn,apj]{emulateapj}

\usepackage{natbib}
\usepackage{graphicx}
\usepackage{amsmath,amssymb}
\usepackage{graphicx}
\usepackage{txfonts}

\usepackage{ulem}
\usepackage{natbib}

\newcommand{\dt}[1] {\frac{\partial{#1}}{\partial t}}
\newcommand{\dr}[1] {\frac{\partial{#1}}{\partial r}}

\newcommand{\dlog}[1] {\frac{1}{{#1}}\frac{d{#1}}{d r}}
\newcommand{\quat}{\frac{1}{4}}

\begin{document}
\title {On the origin of the $1/f$ spectrum in the solar wind magnetic field}
\author{Andrea~Verdini}
\email{verdini@oma.be}
\affil{Solar-Terrestrial Center of Excellence - SIDC, Royal
Observatory of Belgium, Bruxelles}

\author{Roland~Grappin}
\email{Roland.Grappin@obspm.fr}
\affil{LUTH, Observatoire de Paris, CNRS, Universit\'e Paris-Diderot, 
and LPP, Ecole Polytechnique, Palaiseau}

\author{Rui~Pinto}
\email{rui.pinto@cea.fr}
\affil{Laboratoire AIM Paris-Saclay, CEA/Irfu, and Universit\'e
Paris-Diderot CNRS/INSU, Gis-sur-Yvette, France}

\author{Marco~Velli}
\email{mvelli@jpl.nasa.gov}
\affil{Jet Propulsion Laboratory, California Institute of Technology, Pasadena}

 \date{\today}
 
\begin{abstract}
We present a mechanism for the formation of the low frequency $1/f$ magnetic
spectrum based on numerical solutions of a shell reduced-MHD model of the turbulent dynamics inside the sub-Alfv\'enic solar wind.
We assign reasonably realistic
profiles to the wind speed and the density along the radial direction, and a radial magnetic field.
Alfv\'en waves of short periodicity (600 s) are injected at the base
of the chromosphere, penetrate into the corona and are partially reflected, thus
triggering a turbulent cascade. 
The cascade is strong for the reflected wave while it
is weak for the outward propagating waves. Reflection at the transition region
recycles the strong turbulent spectrum into the outward weak spectrum, which is
advected beyond the Alfv\'enic critical point without substantial evolution. 
There, the magnetic field has a perpendicular power-law spectrum
with slope close to the Kolmogorov $-5/3$. 
The parallel spectrum is inherited from the frequency spectrum of
large (perpendicular) eddies. The shape is
a double power-law with slopes of  $\simeq \, -1$ and $-2$ at low and high
frequencies respectively, the position of the break depending on the injected spectrum. 
We suggest that the double power-law spectrum measured by Helios at 0.3~AU,
where the average magnetic field is not aligned with the radial (contrary
to our assumptions) results from the combination of such different spectral
slopes. At low frequency the parallel spectrum dominates with its characteristic
$1/f$ shape, while at higher frequencies its steep spectral slope ($-2$) is
masked by the more energetic perpendicular spectrum (slope $-5/3$). 
\end{abstract}
\section{Introduction}
At heliocentric distances of about 0.3 AU the 
magnetic field spectrum in the fast streams of the solar wind has a form of a
double power-law with a break at about $f=5~10^{-3}~\mathrm{Hz}$ and
a slope of approximately $-1$ and $-5/3$ at lower and higher frequencies
respectively
\citep{Bavassano_al_1982,Denskat_Neubauer_1983,Bruno_Carbone_2005}. 
Fluctuations in these two ranges have a different evolution with distance
\citep{Bavassano_al_1982, Marsch_Tu_1990,Roberts_1992}. The
low frequency part approximately follows the WKB behavior, $E\propto r^{-3}$,
dictated by the solar wind expansion. The high frequency part instead
decreases much sharply, its energy content being depleted by the
turbulent cascade,
and maintains approximately the same spectral slope. Therefore, the
frequency break shifts to lower and lower frequency as the heliocentric
distance increases, the $1/f$ spectrum occupying the range
$3~10^{-6}~\mathrm{Hz}\lesssim f \lesssim 8~10^{-5}~\mathrm{Hz}$ 
at 1~AU \citep{Matthaeus_Goldstein_1986}. 
The energy in the low frequency tail progressively contributes to the turbulent heating of the solar wind. 
The spectral evolution can be understood as a competition between the
expansion timescale and the cascade timescale, regulating the decay of
energy respectively at small and at large scales \citep{Tu_al_1984}, but the origin of the $1/f$ spectrum is still a matter of debate.

There are some indications that it may have a genuine solar
origin, reflecting the emergence, cancellation, and sinking of the
magnetic field at the photosphere: spectra built from magnetogram data at
low-intermediate latitudes show a $1/f$ slope at low wavenumbers,
which also correlate to the spectrum of density
fluctuations in the heliosphere \citep{Matthaeus_al_2007}. 
In this view, the formation mechanism relies on magnetic reconnection, occurring
over a hierarchy of scales as a stochastic process with some self-similar
properties. Reconnection, this time in the corona, 
has also been invoked as the underlying process, since 
the timescale associated to the restructuring of coronal magnetic field is
about 1 day, i.e. in the range of observed frequencies \citep{Close_al_2004}. 
Recently the formation of a $1/f$ spectrum has been observed in homogeneous 
incompressible MHD simulations of very long duration, probably
originating from an inverse cascade not associated with well defined invariants \citep{Dmitruk_Matthaeus_2007,Dmitruk_al_2011}. 
Other ideas rely on large
scale properties of the corona and solar wind: reflection was suggested as a possible mechanism to obtain
a $k^{-1}$ spectrum from an isotropic cascade 
of Alfv\'en waves waves coming from the Sun in the expanding solar wind \citep{Velli_al_1989}. 
The linear evolution of Alfv\'enic pulses excited in the corona
leads to a signal (outside the Alfv\'enic critical point) which has 
periodicities of about 15-30 minutes (ringing of the corona, 
\citealt{Hollweg_Isenberg_2007}).
Though this period appears too low to account for the entire range of observed
frequencies, as suggested by the authors, the ringing of the corona could play a role in
the formation of the $1/f$ spectrum.

In this Letter, we combine the latter two ideas, the ringing of the corona and nonlinear
interactions with reflected waves, to study the formation of the magnetic field spectrum advected
by the solar wind. To this aim, we use a Shell Reduced MHD model \citep{Verdini_al_2009} 
to follow the onset of turbulence resulting from the coupling between Alfv\'en waves
propagating in the chromosphere, corona, and sub-Alfv\'enic solar wind. 
The solar wind profiles (Alfv\'en speed, velocity and density profiles) are given steady 
numerical solutions of the slow wind model by \citet{Pinto_al_2009}. Such slow
wind solutions are chosen to illustrate the mechanism responsible for the
formation of the $1/f$ spectrum, fast wind solutions will be considered in a forthcoming paper.

In homogeneous turbulence with equal energy in both Alfv\'en species the
forcing controls the development of a weak (or strong) cascade, depending on whether the parallel Alfv\'en time 
is smaller (or not) than the nonlinear time. The energy density scales with
perpendicular wave number as $k_\bot^{-2}$ or $k_\bot^{-5/3}$ (e.g.
\citealt{Verdini_Grappin_2012}) for both species.
However, in the inhomogeneous stratified open corona and solar wind, the two
Alfv\'en wave species can in principle have very different amplitudes. We will
see that they follow different cascade regimes (weak/strong) at the same time. This is the key of the results obtained in the present Letter.
We show that for realistic conditions
a $1/f$ magnetic spectrum forms as a consequence of the strong turbulent
cascade of reflected/inward propagating waves, which reflect back (ringing) at
the transition region (TR) and are advected outside the Alfv\'enic critical
point without substantial modification, since the outward propagating waves
experience a weak cascade.
\section{Equations and parameters}
\begin{figure}[t]
\begin{center}
\includegraphics [width=\linewidth]{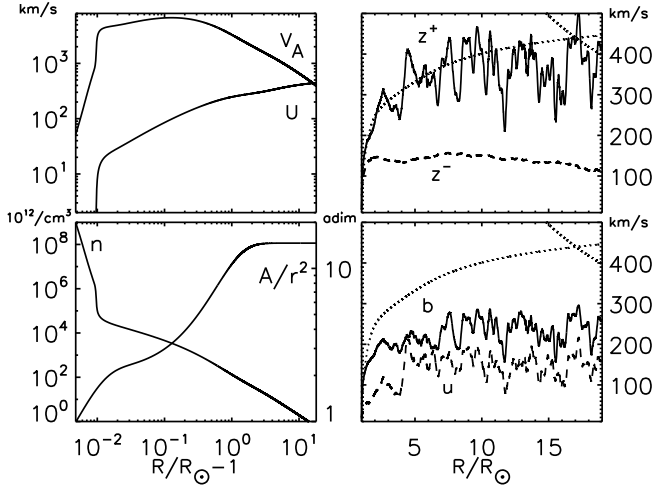}
\caption{\textit{Left panels}: solar wind model. Profiles of the Alfv\'en and
wind speed (top) and of the density and over-expansion ${\cal F}=A(r)/r^2$ (bottom). 
\textit{Right panels}: snapshots of the amplitude of the Els\"asser variables (top) and of the velocity and magnetic
fluctuations (bottom). The Alfv\'en and wind
speed are also overplotted as dotted lines.}
\label{fig1}
\end{center}
\end{figure}

The model equations are obtained from the MHD equations 
by separating the large-scale \textit{stationary} fields and small-scale fluctuating
fields \citep{HO_1980, Velli_1993}. Among the large-scale
fields, the wind speed ($U$) and density
($\rho$) are the solution of a 1D Hydro-dynamic solar wind model
\citep{Pinto_al_2009,Grappin_al_2010} with specified heating functions and
assigned flux tube expansion ($A={\cal F}(r)r^2$). Combining $\rho$ with the
magnetic field strength $B_\odot$ at the base of the fluxtube, and requiring
$BA=const$ one obtains the Alfv\'en speed
($V_a=B/\sqrt{4\pi\rho}$). 
The profiles of $U,~V_a,~\rho$, and ${\cal F}$ depend only on the radial distance $r=R/R_\odot$ and are plotted in the left panels of Figure~\ref{fig1}: 
the bottom and top boundaries are at $r_{bot}=1.004$ and
$r_{top}=19$, the Alfv\'enic critical point is found at
$r_A\approx17$ by choosing $B_\odot=10~\mathrm{G}$.
The small-scale \textit{incompressible} velocity ($u$) and magnetic ($b$) fluctuations are orthogonal to the radial $B$ and can be expressed through the usual Els\"{a}sser variables $z^\pm=u\mp b/\sqrt{4\pi\rho}$.
We further simplify the equations by replacing the nonlinear terms and the
pressure term at each position by a 2D shell
model, so that the final equations read:
{\setlength\arraycolsep{2pt}\begin{eqnarray}
\dt{z^\pm_n} &+& (U\pm V_a)\dr{z^\pm_n}  
- \quat(U\mp V_a)\left(\dlog{\rho}\right) z^\pm_n  \nonumber\\
&+& \quat(U\mp V_a)\left(\dlog{\rho} +2\dlog{A}\right)z^\mp_n  = T_{npq}^\pm - \nu
k_n^2 z^\pm
\label{eq:wave}
\end{eqnarray}}
where we take equal viscosity and resistivity ($\nu$) and $T_{npq}^\pm$ denotes the nonlinear interactions.
The index $n$ labels 
modes having perpendicular wavenumber 
$k_n=k_02^n$, which define the radius of concentric shells filling
the (perpendicular) Fourier space. The largest transverse
scale follows the flux tube
expansion $k_0(r) = k_{0\odot}/\sqrt{A(r)}$, wavevectors are 
given by $k_\bot\equiv k_n(r)=k_0(r)2^n$, so that our Fourier space 
shrinks with increasing $r$. Two complex fields
are assigned to each shell, $z^\pm_n(r,t)\equiv z^\pm(k_n,r,t)$. They have the dimension of a
velocity and $|z_n^\pm|^2/2$ are the respective 
energies per unit mass in the shell $n$.  
We recall that nonlinear interactions are local in Fourier space, $T_{npq}\propto \Sigma_{p,q} k_n z^\pm_p z^\mp_q$, with $p\sim q\sim n$ (the explicit expression and coefficients for the 2D shell model may be found in
\citealt{Giuliani_Carbone_1998}). 

Open boundaries are used at $r_{bot}$ and $r_{top}$.
The free parameters are the
input wave amplitude $z^+_\odot$, the perpendicular injection scales
$k_{0\odot}$ (the forcing being imposed on $1,2,4~k_{0\odot}$), and the
correlation time of the forcing signal $T_f$.
We choose a \textit{strong} forcing, i.e., $T_f \simeq t_{NL}^0$ by assigning:
$z^+_\odot=10~\mathrm{km/s}$,
$k_{0\odot}=2\pi/34000~\mathrm{km^{-1}}$, and $T_f=600~\mathrm{s}$. 
With these parameters
$t_{NL}^0=1/k_0^\odot z^+_\odot\simeq 500~\mathrm{s} \simeq T_f$.

Decreasing the free parameter $B_\odot$ shifts the profile of
$V_a$ in Figure~\ref{fig1}, to lower values, thus decreasing $r_A$ and resizing the regions where
one linear term dominates over the other linear terms
(propagation, WKB, and reflection, respectively
the II, III, and IV terms on the LHS of Eq.~\ref{eq:wave}).

Starting from the solutions $z_n^\pm(r,t)$ of Eq.~\ref{eq:wave} we define frequency and wave number spectra.
We denote by
$\hat z_n^\pm(r,f)$ their Fourier transform with respect to time.
From this we define successively 
the spectra $E_n^\pm(f)=E^\pm(n,f)$
and the associated reduced spectra at each radial distance $r$:
\begin{eqnarray}
E^\pm_n(f) &=&  |\hat z_n^\pm(r,f)|^2 /k_\perp
\label{cuts}\\
E_f^\pm(f) &=& \textstyle{\int} E_n^\pm dk_\perp
\label{1dpar}\\ 
E_\perp^\pm(k_\perp) &=&  \textstyle{\int} E_n^\pm df
\label{1dperp}
\end{eqnarray}
where we have omitted the dependence on $r$.
Equation~(\ref{cuts}) is the frequency spectrum of a given perpendicular
mode which yields the total energy $E^\pm(r)=\sum_n |z_n^\pm|^2 = \int E_n^\pm dk_\perp df$. 
Equations~(\ref{1dpar}-\ref{1dperp})
define reduced spectra, which accordingly yield
$\int E_f^\pm df=\int E_\perp^\pm dk_\perp=E^\pm$ (hereafter we drop the subscripts $f,~k_\bot$ when the dependence is explicit). 

The correlation time $t^\pm_{cor}(r,~k_\perp)$ is defined as
the full-width-half-maximum (FWHM) of the autocorrelation function
$AC(z^\pm)$ computed at each position $r$, 
while the nonlinear time $t^\pm_{NL}(r,~k_\perp)$ is the time average of the
eddy turnover times built from the Els\"asser fields \citep{Dobrowolny_al_1980}:
\begin{eqnarray}
t^\pm_{cor}(k_\perp) &=&  \textrm{FWHM}[AC(z^\pm)_t] \label{tcor}\\
t^\pm_{NL}(k_\perp) &=& \langle1/(k_\perp |z^\mp(k_\perp)|)\rangle \label{tnl}
\end{eqnarray}
Again we omitted the dependence on $r$, 
$AC(z)=\int z(t)z(t-t') dt'$ is the autocorrelation function and
$\langle...\rangle$ stands for a time average.
The amplitudes of the fluctuations are computed by integrating along the
perpendicular wavenumbers $z^\pm=\sqrt{\Sigma_n |z_n(r,t)|^2}$.
The same definitions hold for $u$ and $b$, the latter will
be expressed in velocity units from now on (i.e. $b\rightarrow
b/\sqrt{4\pi\rho}$).
The equality of the two timescales defines the so called critical balance (CB)
condition:
\begin{eqnarray}
t^\pm_{cor}(k)= t^\pm_{NL}(k), 
\label{eq:CB}
\end{eqnarray}
that is supposed to hold for strong turbulence. 
Since the correlation time is the inverse of the width of the frequency
spectrum, we can define an equivalent CB width $\Delta^\pm_{CB}=1/t_{NL}^\pm$.\\
According to anisotropic turbulence theories
\citep{GS_1995}, this width constrains the frequency spectrum, since at a given perpendicular scale $k_\bot$ there is little 
energy for $f>\Delta_{CB}$, while spectra are flat for $f<\Delta_{CB}$.
\section{Results}
Figure~\ref{fig1} shows a snapshot of the amplitude of Alfv\'en waves
(top right) and of the kinetic and
magnetic fluctuations (bottom right). 
In the corona 
$z^+\simeq 1.5 b$, $b \simeq 1.5 u$. The profile of $z^+$ 
follows the wind speed profile (background dotted line) and is much larger the
inward/reflected wave $z^-$. $z^-$ is a smooth function of distance, while $z^+$ displays small scale parallel
structure, seen also in the radial profiles of $u$ and $b$.
\begin{figure}[t]
\begin{center}
\includegraphics [width=\linewidth]{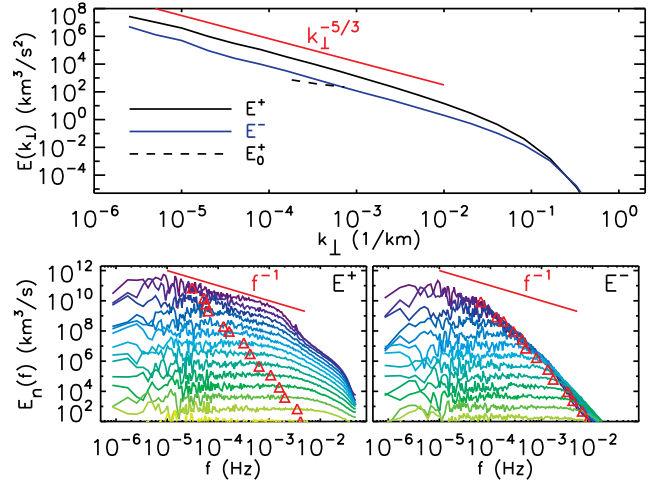}
\caption{\textit{Top}. Perpendicular spectra $E^\pm(k_\bot)$ at $r_{top}$.
The input spectrum $E^+_0$ at $r_{bot}$ is also shown (dashed line).
\textit{Bottom}. Frequency spectra $E_n^\pm$ at
$r_{top}$ for perpendicular wavenumbers $k_0 2^n$ (with $n$
increasing from top to bottom). Symbols mark the CB width 
$\Delta^\pm_{CB}$ (see text). See the electronic edition of the Journal for a color version of this figure.}
\label{fig2}
\end{center}
\end{figure}

The spectral densities vs perpendicular wave number and frequency at $r_{top}=19$ are
summarized in Figure~\ref{fig2}. In the top panel the time-averaged
perpendicular spectra $E^\pm(k_\bot)$ are plotted along with the input
spectrum at the base of the chromosphere ($E_0^+$).
Both perpendicular spectra show a well developed power-law that extends about
two decades with a slope $-5/3$. 
The ratio $E^+_\bot/E^-_\bot$ is about constant in the inertial range.
The input perpendicular spectrum $E^+_0$ (dashed line) appears in the middle of the inertial range of the spectra at $r_{top}$, as the flux tube expansion has strongly expanded the perpendicular scales between the surface and $r_{top}$ (by a factor $100$).

In the bottom panels of Figure~\ref{fig2} the frequency
spectra, $E_n^\pm(f)$, are plotted for each perpendicular
mode $n$. Symbols mark the equivalent CB width $\Delta^\pm_{CB}$. 
The $E^-$ spectra (right) are very well bounded by the CB
condition, falling off sharply in the weak turbulence range
($f>\Delta^-_{CB}$), and having flat spectra in the strong turbulence range
($f<\Delta^-_{CB}$), with few exceptions for large-scale eddies ($n=0,1,2$).
On the contrary $E_n^+$ has a substantial over-excitation of high
frequencies at large perpendicular scales and in the whole weak range,
which reflects the fine-scale (parallel) spatial structure of $z^+$ seen
in Figure~\ref{fig1}.
This over-excitation has an approximate slope $1/f$ and extends from the
CB boundary to about $1/T_f \simeq 2 \ 10^{-3}~\mathrm{Hz}$. 

The slope of the frequency spectrum has been found to depend on turbulence forcing strength $T_f/t_{NL}^0$ in homogenous \textit{shell} RMHD
simulations \citep{Verdini_Grappin_2012}, and to extend
beyond forced scales. 
The situation is different here because we cannot really control the
forcing turbulence strength: the nonlinear time for $z^+$ depends on the
reflected amplitude $z^-$, an output of the simulation. 
An \textit{a posteriori} estimate yields
a turbulence strength $t^+_{cor}/t_{NL}^+<1$ so that $z^+$ is indirectly
forced in the weak regime.
Coming back to the strong turbulence range, one can see that large
eddies have an approximate slope of $1/f$ in both $E^-$
and $E^+$: these low-frequency fluctuations
are \textit{linearly} coupled by density gradients (reflection) that force the
spectra to resemble each other.

\begin{figure}[t]
\begin{center}
\includegraphics [width=\linewidth]{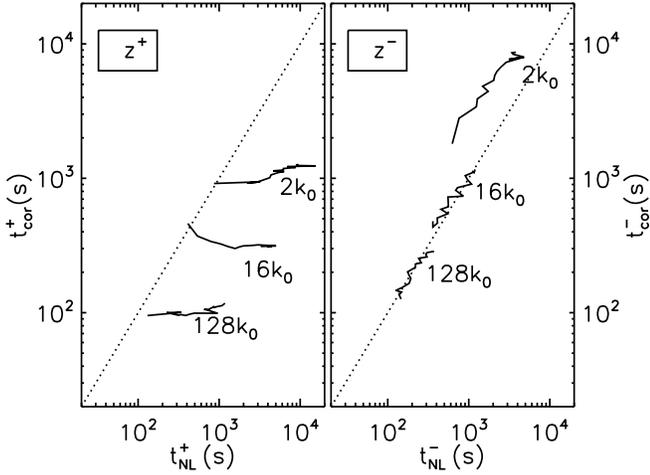}
\caption{Correlation time $t_{cor}^\pm$ versus nonlinear time $t_{NL}^\pm$ 
for different heliocentric distance (running from $2R_\odot$ to $19R_\odot$ from left to right on each curve) at three perpendicular wavenumbers
$2k_\bot^0,~8k_\bot^0,~128k_\bot^0$. The dotted line is the CB condition
Eq.~\ref{eq:CB}.}
\label{fig3}
\end{center}
\end{figure}
The turbulence regime at different heliocentric distances can be identified in Figure~\ref{fig3}, where we
plot the correlation time vs the nonlinear time for three 
wavenumbers ($2k_0,~8k_0,~128k_0$) at heliocentric distances,
running from left
($r=2$) to right ($r=19$) along the solid lines. The critical balance condition Eq.~\ref{eq:CB} is drawn as a
dotted line and separates the strong and weak regimes (above and below respectively). 
The reflected wave $z^-$ is always in a strong turbulence regime and follows the
CB condition at any position, except for a small offset at low wavenumbers
due to the linear coupling (reflection) at large scale. 
On the contrary the correlation time of $z^+$ is
almost horizontal, i.e. almost independent of the nonlinear time and of
distance, and lies entirely in the weak turbulence regime.
Since $t_{cor}$ is the inverse of the width of the frequency
spectrum, one can conclude that 
the $E^+$ spectrum does not change much as it
propagates outward, only becoming a bit narrower at large scales. 
Instead, the $E^-$ spectrum widens while propagating backward from $r_A$,
according to CB. 
Once it arrives at the TR it experiences strong reflection and feeds the $E^+$
spectrum, this can be argued by noting that $t^+_{cor}\lesssim t^-_{cor}$ at $r=2$ (the low end of the curves).
At larger distances $t^+_{cor}<<t^-_{cor}$,
showing that $E_n^+$ is wider than $E^-_n$ in the
whole corona, the difference owing to its high-frequency over-excitation.

\begin{figure}[t]
\begin{center}
\includegraphics [width=\linewidth]{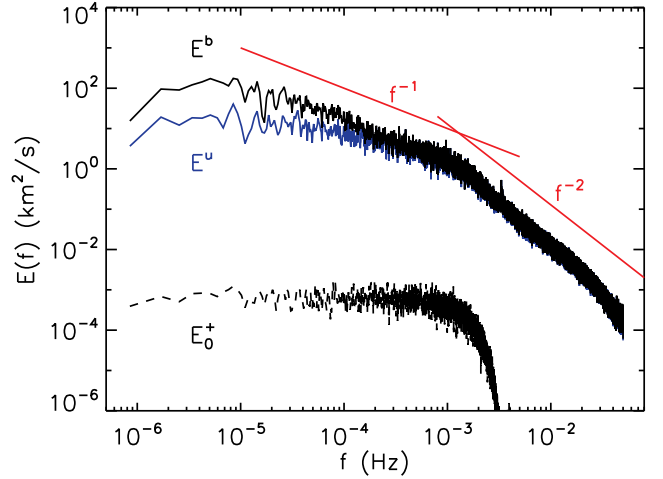}
\caption{Reduced frequency spectra for $E^{u,b}$ at $r_{top}$ (solid
gray and black line respectively) and input spectrum $E^+_0$ at $r_{bot}$
(dashed line). See the electronic edition of the Journal for a color version 
of this figure.}
\label{fig4}
\end{center}
\end{figure}

The $1/f$ spectrum appearing at large perpendicular scales in
$E^+$ is thus made up of two parts, each occupying about one decade from
$f\simeq10^{-5}~\mathrm{Hz}$, which originate from two different mechanisms:
linear coupling with $E^-$ at low frequencies (the strong turbulence
range), and weak turbulent cascade at intermediate frequencies.
This is summarized in Figure~\ref{fig4} where the reduced frequency spectra $E^{u,b}(f)$ of kinetic
and magnetic energy at $r=r_{top}$ are shown,
(along with the input spectrum at the base of the chromosphere $E^+_{0}$). The spectrum of $E^b$ shows the $1/f$ slope inherited
from the large scale eddies in $E^+$, in which one can recognize the above two
components, and a break appears at about $f\simeq10^{-3}~\mathrm{Hz}$ where the slope switches to
$-2$. The break coincides with the width of the forcing spectrum 
($1/T_f\simeq 2~10^{-3}~\mathrm{Hz}$) and the $-2$ slope is consistent with
the fall-off at high frequencies in $E_n^b\simeq E_n^+$.
Note that linear propagation alone would lead to a
magnetic spectrum with slope $f^{-1/2}$ (not shown), thus nonlinear
interactions are fundamental to obtain the $1/f$ slope.
The spectrum of $E^u$ is practically identical to $E^b$ for
$f\gtrsim 10^{-4}~\mathrm{Hz}$ due to the same over-excitation 
in the weak regime.
For lower frequencies instead the slope of $E^u$ is flatter, having a value
of about $-1/2$. 
\section{Discussion}
Why the $1/f$ extends down to low frequencies only in $E^b$
can be understood by examining the wave reflection at the
TR, where density gradients are higher.
For frequencies $f\lesssim \mathrm{max}[|dV_a/dr|]\simeq 2~\mathrm{Hz}$, 
one can write $z^-=-(1-\epsilon)z^+$ where $\epsilon =V_a^{chrom}/V_a^{cor}<<1$
is the Alfv\'en speed contrast between the chromosphere and corona
\citep{Hollweg_1984,VanBallegooijen_al_2011,Verdini_al_2012}. At the TR,
$b\simeq z^\pm$ and $u<<z^\pm$, thus reflection transfers the
low-frequency large-eddy spectrum of $E^-$ to $E^b$ but not to $E^u$.
In reality reflection occurs in the whole low corona (for $r\lesssim 3~R_\odot$
at frequencies $f<5~10^{-3}~\mathrm{Hz}$) and is a continuous process.
Moreover, reflected waves are made of two components: a classical one,
propagating backward in the region $U<V_a$ and subject to ringing in the low
corona, and an anomalous
component that instead propagates outward with the mother wave
\citep{Velli_al_1989,Hollweg_Isenberg_2007,Verdini_al_2009}. 
Nonlinear interactions are not limited to the duration of the encounter of
colliding $z^\pm$ wavepackets, but part of them occur on
the common outward propagation path. This is why linear and also nonlinear
couplings affect the low tail of the $1/f$ spectrum.

Bearing in mind that
$z^-$ is entirely generated by reflection inside the solar atmosphere, we
examine now what happens if we increase the input turbulence strength (i.e the
control parameter $T_f/t_{NL}^0$) by either
increasing the correlation time $T_f$ or decreasing the input nonlinear time
$t^0_{NL}=1/k_{0\odot}z^+_\odot$.
By imposing a shorter input nonlinear time, no matter if through $z^+_\odot$ or
$k_{0\odot}$, we increase the strength of the cascade for $z^-$. The
strength of the $z^+$ cascade will increase only slightly, since the reflected
wave will be damped more strongly, yielding approximately the same $t_{NL}^+$. The spectrum of $E^+$
will thus be affected only slightly, except that now nonlinear coupling
will dominate over reflection in the low corona, eroding the very low frequency tail of the $1/f$ spectrum in $E^b$.
If instead we choose a longer 
correlation time at input, we increase the amount of reflected waves thus
also increasing the strength of the cascade for $z^+$. This time the
high-frequency tail of the $1/f$ spectrum will be eroded, since a
stronger cascade reduces the energy in the over-excited weak-turbulence regime of $E^+$.
Finally, one can vary the importance of linear/nonlinear coupling through the 
other free parameter $B_\odot$. Taking a smaller chromospheric magnetic field
will lower the Alfv\'enic critical point, reducing the importance of linear
terms compared to the nonlinear ones in Eq.~\ref{eq:wave}. 
The changes in the $E^b$ spectrum are similar to those discussed above for a
shorter $t^0_{NL}$. Note however that halving $B_\odot$ moves $r_A$ to a
distance of $8~R_\odot$, so that one is limited to small variations of this
parameter.
A last remark concerns the non-local couplings in the Fourier space that
are neglected in the shell model employed in this work. In principle they could
change the perpendicular spectra and the strength of the cascade. However, our understanding of the process of formation 
of the $1/f$ spectrum is that it comes from the different nature of the two direct $z^-$ and $z^+$ 
cascades, not because of an inverse cascade as in the dynamo process.
We expect these direct cascades not to change when including nonlocal interactions.

The formation of the $1/f$ spectrum by nonlinear and linear coupling in
the sub-Alfv\'enic solar wind appears to be quite solid, however its relation
to the interplanetary spectrum observed at $0.3$~AU is not straightforward.
Beyond $r_A$ the solar wind expansion causes the rotation of the magnetic field, a slower production of reflected waves, and a decrease of wave-numbers perpendicular to the radial direction \citep{Grappin_al_1993}. 
The latter two actually suggest a freezing of the spectra. On the contrary the magnetic field rotation causes
instabilities and wave coupling/decay that could modify the spectrum. 
Numerical simulations of MHD equations in 1D and 2D incorporating the
effect of an expanding solar wind \citep{Grappin_al_1993,Grappin_Velli_1996}
suggest that the decay and instability rates are delayed with respect to
the non-expanding case, so that a freezing of the advected spectra may be a good approximation, at least until $0.3$~AU.
If one assumes no further evolution, the double-power-law
measured by Helios at 0.3 AU, may result from the combination of the parallel
and perpendicular reduced spectra.
At low frequency the parallel spectrum dominates with its characteristic slope $-1$, while at higher frequencies its steep spectral slope ($-2$) is masked by
the more energetic perpendicular spectrum (with a $-5/3$ slope). Our theory
might well be tested by the measurements of Solar Probe Plus.\\

\acknowledgments
\textit{Acknowledgments} We thank E. Buchlin for useful suggestions while implementing the background solar wind in the numerical code.
A.V. acknowledges support from the Belgian Federal Science Policy Office through the ESA-PRODEX program.
The research described in this paper was carried out in part at the Jet
Propulsion Laboratory, California Institute of Technology, under a contract
with the National Aeronautics and Space Administration.

\end{document}